\documentclass[final]{svjour2}
\usepackage[dvips]{graphicx}
\usepackage{rotating}
\usepackage{amssymb}
\usepackage{mathptmx}
\usepackage[numbers]{natbib}
\makeatletter
\journalname{Journal of Low Temperature Physics}
\bibpunct{}{}{,}{s}{}{,}

\def\up{\uparrow}
\def\down{\downarrow}
\def\mbf#1{\mathbf#1}
\def\lesssim{\ \raise.3ex\hbox{$<$}\kern-0.8em\lower.7ex\hbox{$\sim$}\ }
\def\gesim{\ \raise.3ex\hbox{$>$}\kern-0.8em\lower.7ex\hbox{$\sim$}\ }

\begin{document}

\newcommand{\hdblarrow}{H\makebox[0.9ex][l]{$\downdownarrows$}-}
\title{Spin Susceptibility and Strong Coupling Effects in an Ultracold Fermi Gas }

\author{H. Tajima \and R. Hanai \and R. Watanabe \and Y. Ohashi}

\institute{Department of Physics, Faculty of Science and Technology, Keio University,\\ 3-14-1, Hiyoshi, Kohoku-ku, Yokohama 223-8522, Japan\\
Tel.:+81-45-566-1454 Fax:+81-45-566-1672
\email{htajima@rk.phys.keio.ac.jp}
%\\2: XXXXXXXXX\\ 3: YYYYYYYYYYY
}

\date{6.30.2013}

\maketitle

\keywords{Fermi superfluid, Spin-gap phenomenon, BCS-BEC crossover}

\begin{abstract}
We investigate magnetic properties and strong coupling corrections in the BCS (Bardeen-Cooper-Schrieffer)-BEC (Bose-Einstein condensation) crossover regime of an ultracold Fermi gas. Within the framework of an extended $T$-matrix theory, we calculate the spin susceptibility $\chi$ above the superfluid phase transition temperature $T_{\rm c}$. In the crossover region, the formation of preformed Cooper pairs is shown to cause a non-monotonic temperature dependence of $\chi$, which is similar to the so-called spin-gap phenomenon observed in the under-doped regime of high-$T_{c}$ cuprates. From this behavior of $\chi$, we determine the spin-gap temperature as the temperature at which $\chi$ takes a maximum value, in the BCS-BEC crossover region. Since the spin susceptibility is sensitive to the formation of singlet Cooper pairs, our results would be useful in considering the temperature region where pairing fluctuations are important in the BCS-BEC crossover regime of an ultracold Fermi gas. 

PACS numbers: 03.75.Hh, 05.30.Fk, 67.85.Lm.

\end{abstract}
%%%%%%%%%%%%%%%%%%%%%%%%%%%%%%%%%%%%%%%%%%%%%%%%%%%%%%%%%%%%%%%%%%%%%%%%%%%%%%%%%%%%
\section{Introduction}
Recently, the ultracold Fermi gas has attracted much attention as a useful quantum system to study many-body physics in a strongly interacting fermion system\cite{sgiorgini08, ibloch08}. In particular, using a tunable pairing interaction associated with a Feshbach resonance\cite{cchin10}, we can now study superfluid properties from the weak-coupling BCS regime to the strong-coupling BEC limit in a unified manner\cite{sgiorgini08,ibloch08,yohashi03}. In the intermediate coupling regime, which is referred to as the BCS-BEC crossover region in the literature, physical properties of a system are dominated by pairing fluctuations. Thus, in this regime, the possibility of the so-called pseudogap phenomenon has been discussed\cite{stsuchiya09,Chen2,Magierski,rwatanabe10,Hu,Su,Perali,Mueller}, where preformed Cooper pairs cause a gap-like structure in the single-particle excitation spectrum even in the normal state.
\par
However, the existence of the pseudogap in ultracold Fermi gases is still in debate. The photoemission-type experiment on a $^{40}$K Fermi gas has observed anomalous single-particle excitation spectra in the BCS-BEC crossover region\cite{jtstewart08,jpgaebler10}, which supports the pseudogap scenario\cite{stsuchiya09,Chen2,Magierski,rwatanabe10,Hu,Su,Perali,Mueller}. On the other hand,  the local pressure measurement on a $^6$Li Fermi gas denies the pseudogap, where the experimental data are reported to be simply explained by the normal Fermi liquid theory\cite{nascimbene11}. Thus, although it has been pointed out that the latter experimental result can be also explained by the pseudogap scenario\cite{rwatanabe12}, further studies are necessary to resolve this important problem.
\par
In considering this many-body problem, extensive work done in high-$T_{\rm c}$ cuprates would be helpful. In this strongly correlated electron system, the pseudogapped density of states has been observed in the underdoped regime\cite{chrenner98}, although the origin of this phenomenon still remains to be solved, due to the complexity of this system. In addition to this, in the underdoped regime, the so-called spin-gap phenomenon has been observed, where the uniform susceptibility, as well as the nuclear magnetic relaxation rate (NMR-$T_1^{-1}$), exhibit anomalous temperature dependences\cite{yyoshinari90}. In this regard, we note that, when the pseudogap really exists in the BCS-BEC crossover regime of cold Fermi gases, since the spin susceptibility $\chi$ is deeply related to the density of states at the Fermi level, $\chi$ would be suppressed by this phenomenon. Furthermore, the pseudogap in cold Fermi gases originates from pairing fluctuations being accompanied by preformed Cooper pairs, these singlet pairs would also suppress the spin susceptibility. Thus, as a probe to examine the pseudogap phenomenon in cold Fermi gases, the uniform spin susceptibility may be useful. We briefly note that the observation of this quantity has recently become possible in cold Fermi gases\cite{Sanner,Sommer}.
\par
In this paper, we theoretically investigate magnetic properties of an ultracold Fermi gas in the normal state. In considering this problem, we recall that the strong-coupling coupling theory developed by Nozi\`eres and Schmitt-Rink\cite{NSR,SadeMelo}, as well as the ordinary $T$-matrix theory\cite{stsuchiya09,rwatanabe10,Pieri,Stajic}, that have been extensively used to clarify various BCS-BEC crossover physics in cold Fermi gases, unphysically give {\it negative} susceptibility in the crossover region\cite{Liu,Parish,tkashimura12}. This difficulty has been , however, recently overcome by properly including higher order fluctuation effects beyond the $T$-matrix level\cite{tkashimura12}. In this paper, we also employ this extended $T$-matrix theory to calculate the uniform (pseudo)spin susceptibility $\chi$. We show how this quantity is affected by pairing fluctuations in the BCS-BEC crossover region. We also introduce the spin-gap temperature as the temperature below which $\chi$ is suppressed by the preformed pair formation. In this paper, we take $\hbar=k_{\rm B}=1$, and the system volume is taken to be unity, for simplicity.

%%%%%%%%%%%%%%%%%%%%%%%%%%%%%%%%%
\begin{figure}
\begin{center}
\includegraphics[width=8cm]{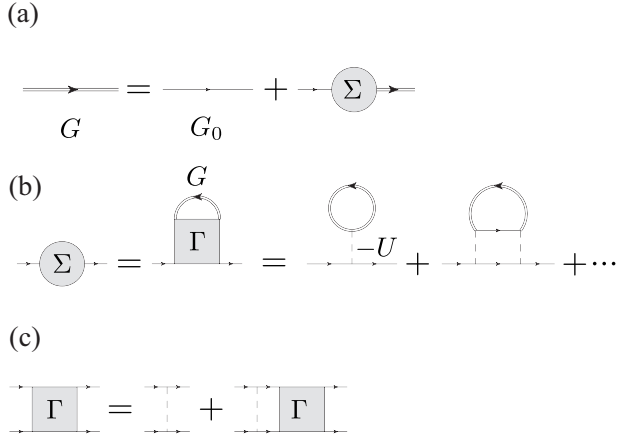}
\end{center}
\caption{(a) Single-particle Green's function $G_{\sigma}(\mbf{p},i\omega_{n})$ (double solid line). The single solid line describes the Green's function $G_{0\sigma}(\mbf{p},i\omega_{n})$ for a free Fermi gas. (b) Self-energy correction $\Sigma_{\sigma}(\mbf{p},i\omega_{n})$ in the extended $T$-matrix theory. The dashed line describes the pairing interaction $-U$. (c) Particle-Particle scattering matrix $\Gamma(\mbf{q},i\nu_{n})$. We note that the bare Green's function $G_0$ is still used in $\Gamma(\mbf{q},i\nu_{n})$ in the extended $T$-matrix theory\cite{tkashimura12}.}
\label{fig1}
\end{figure}
%%%%%%%%%%%%%%%%%%%%%%%%%%%%%%%%%

%%%%%%%%%%%%%%%%%%%%%%%%%%%%%%%%%%%%%%%%%%%%%%%%%%%%%%%%%%%%%%%%%%%%%%%%%%%%%

\par\section{Formulation}
\par
We consider a two-component Fermi gas, described by the BCS Hamiltonian,
\begin{equation}
   H = \sum_{\mbf{p},\sigma}\xi_{\mbf{p},\sigma}c^{\dag}_{\mbf{p},\sigma}c_{\mbf{p},\sigma}-U\sum_{\mbf{p},\mbf{p'},\mbf{q}} c^{\dag}_{\mbf{p}+\mbf{q}/2,\up}c^{\dag}_{-\mbf{p}+\mbf{q}/2,\down}c_{-\mbf{p'}+\mbf{q}/2,\down}c_{\mbf{p'}+\mbf{q}/2,\up}.
   \label{eq.1}
\end{equation}
Here, $c^{\dag}_{\mbf{p},\sigma}$ is the creation operator of a Fermi atom in the hyperfine state described by pseudospin $\sigma=\uparrow,\downarrow$. $\xi_{\mbf{p}\sigma}=p^{2}/(2m)-\mu_{\sigma}$ is the kinetic energy of the $\sigma$-spin component, measured from the chemical potential $\mu_\sigma$ (where $m$ is an atomic mass). When one writes the chemical potential as $\mu_{\sigma}=\mu + \sigma h$, $h$ may be viewed as an effective magnetic field, which we will use in calculating the spin susceptibility $\chi$. When $h>0$, the system has a population imbalance $\Delta N = N_{\up}-N_{\down}$ (where $N_\sigma$ is the total number of Fermi atoms in the $\sigma$-spin component). In Eq. (\ref{eq.1}), $-U (<0)$ is a tunable pairing interaction associated with a Feshbach resonance. For simplicity, we ignore effects of a harmonic trap in this paper. 
\par
As usual, we measure the interaction strength in terms of the $s$-wave scattering length $a_{s}$, given by
\begin{equation}
\frac{4 \pi a_{s}}{m}=\frac{-U}{1-U\sum^{\omega_{c}}_{p}\frac{m}{p^2}},
\end{equation}
where $\omega_{c}$ is a cut-off energy. In this scale, the weak-coupling BCS regime and the strong-coupling BEC regime are given by $(k_{\rm F}a_{s})^{-1} \lesssim -1 $ and $ (k_{\rm F}a_{s})^{-1} \gesim 1$, respectively (where $k_{\rm F}$ is the Fermi momentum). The crossover region is characterized as $-1 \lesssim (k_{\rm F}a_{s})^{-1} \lesssim 1 $.
\par
The uniform spin susceptibility $\chi$ is calculated from
\begin{equation}
\chi=\lim_{h\rightarrow 0}\frac{\Delta N}{h}=\lim_{h\rightarrow 0}\frac{N_{\up}-N_{\down}}{h}.
\label{eq.2}
\end{equation}
In this paper, we numerically evaluate Eq.(\ref{eq.2}), by taking a small but finite value of $h$. The particle number $N_\sigma$ in the $\sigma$-spin component in Eq.(\ref{eq.2}) is calculated from
\begin{equation}
N_{\sigma}=T\sum_{\mbf{p},i\omega_{n}}G_{\sigma}(\mbf{p},i\omega_{n}).
\label{eq.3}
\end{equation}
Here, $G_{\sigma}(\mbf{p},i\omega_{n})$ is the single-particle thermal Green's function, having the form
\begin{equation}
G_{\sigma}(\mbf{p},i\omega_{n}) = \frac{1}{i\omega_{n}-\xi_{\mbf{p},\sigma}-\Sigma_{\sigma}(\mbf{p},i\omega_{n}) },
\label{eq.4}
\end{equation}
where $\omega_n$ is the fermion Matsubara frequency. The self-energy $\Sigma_{\sigma}(\mbf{p},i\omega_{n})$ involves fluctuation corrections to single-particle excitations. In the extended $T$-matrix theory\cite{tkashimura12}, it is diagrammatically given by Fig.\ref{fig1}. Summing up the diagrams in this figure, one has
\begin{equation}
\Sigma_{\sigma}(\mbf{p}, i\omega_{n})=\sum_{\mbf{q},i\nu_{n}}\Gamma(\mbf{q},i\nu_{n})G_{-\sigma}(\mbf{q}-\mbf{p},i\nu_{n}-i\omega_{n}),
\label{eq.5}
\end{equation} 
where $\nu_{n}$ is the Boson Matsubara frequency. The particle-particle scattering matrix $\Gamma(\mbf{q},i\nu_{n})$ has the form,
\begin{equation}
\Gamma(\mbf{q},i\nu_{n})=\frac{-U}{1-U\Pi (\mbf{q},i\nu_{n})},
\label{eq.6}
\end{equation}
where 
\begin{equation}
\Pi (\mbf{q},i\nu_{n})=\sum_{\mbf{p},i\omega_{n}}G_{0\up}(\mbf{p}+\mbf{q}/2,i\nu_{n}+i\omega_{n})G_{0\down}(-\mbf{p}+\mbf{q}/2,-i\omega_{n}),
\label{eq.7}
\end{equation}
is the lowest-order pair correlation function. $G_{0\sigma}(\mbf{p},i\omega_{n})=1/(i\omega_{n}-\xi_{\mbf{p},\sigma})$ is the single-particle Green's function for a free Fermi gas.
\par
\begin{figure}
\begin{center}
\includegraphics[width=10cm]{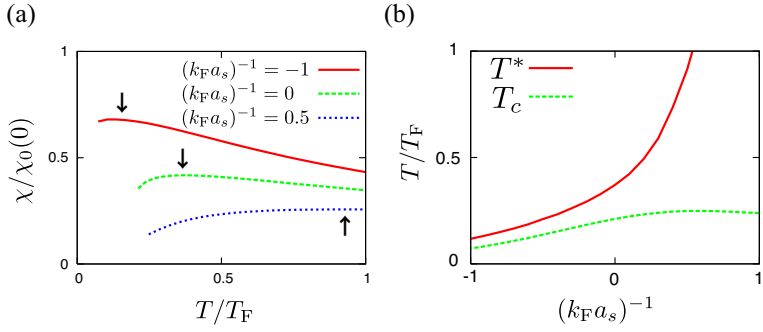}
\end{center}
\caption{(Color online) (a) Calculated spin susceptibility $\chi$ as a function of temperature. $\chi_0(0)$ is the spin susceptibility at $T=0$ for a free Fermi gas, $T_{\rm F}$ is the Fermi temperature. $k_{\rm F}$ is the Fermi momentum. The arrows show the spin-gap temperatures. (b) Spin-gap temperature $T^{*}$ (solid line). The dashes line shows $T_{c}$.}
\label{fig2}
\end{figure}

The superfluid phase transition temperature $T_{\rm c}$ is conveniently determined from the Thouless criterion, 
\begin{equation}
\Gamma^{-1}(\mbf{q}=\mbf{0},i\nu_{n}=0)=0,
\label{eq.8}
\end{equation}
together with the equation for the total number $N$ of Fermi atoms,
\begin{equation}
N=N_{\up}+N_{\down}.
\label{eq.9}
\end{equation}
In solving the coupled equations (\ref{eq.8}) and (\ref{eq.9}), we set $h=0$ ($\mu_\up=\mu_\down$). We show the calculated $T_{\rm c}$ in Fig.\ref{fig2}(b). 
\par
Once $T_{\rm c}$ is determined, we only solve the number equation (\ref{eq.9}) above $T_{\rm c}$ for a small but finite value of $h$, to determine the average chemical potential $\mu\equiv(\mu_\up+\mu_\downarrow)/2$. We then calculate the spin susceptibility in Eq. (\ref{eq.2}). In this procedure, the small value of $h$ is chosen so that the numerator in Eq.(\ref{eq.2}) be proportional to $h$. 
\par
We briefly note that the ordinary (non-selfconsistent) $T$-matrix theory is recovered, when the full Green's function $G_\sigma$ in Fig.\ref{fig1}(a) is simply replaced by the non-interacting one $G_{0\sigma}$. Because of this improvement, the calculated spin susceptibility in the present extended $T$-matrix theory satisfies the required positivity in the whole BCS-BEC crossover region\cite{tkashimura12}.
\par

%%%%%%%%%%%%%%%%%%%%%%%%%%%%%%%%%%%%%%%%
%%%%%%%%%%%%%%%%%%%%%%%%%%%%%%%%%%%%%%%%

%%%%%%%%%%%%%%%%%%%%%%%%%%%%%%%%%%%%%%%%%%%%%%%%%%%%%%%%%%%%%%%%%%%%%%%%%%%%%%%
\par
\section{Spin-gap Phenomenon in the BCS-BEC Crossover Region}
\par
Figure \ref{fig2}(a) shows the temperature dependence of the spin susceptibility $\chi$ in the BCS-BEC crossover region. In the BCS regime ($(k_{\rm F}a_s)^{-1}=-1$), $\chi$ increases with decreasing the temperature when $T/T_{\rm F}\gtrsim 0.12$, which is similar to the temperature dependence of the spin susceptibility $\chi_0$ for a free Fermi gas, given by\cite{Kubo}
\begin{equation}
\chi_0(T)\simeq \chi_0(0)
\Bigl[
1-{\pi^2 \over 12}
\Bigl(
{T \over T_{\rm F}}
\Bigr)^2
\Bigr]~~~(T\ll T_{\rm F}).
\end{equation}
However, in contrast to the free Fermi gas, $\chi$ decreases with decreasing the temperature when $T/T_{\rm F}\lesssim 0.12$. Since the spin degrees of freedom is suppressed by preformed {\it singlet} Cooper pairs, this decrease is considered to originate from pairing fluctuations. Indeed, this low temperature behavior gradually becomes remarkable, as one passes through the BCS-BEC crossover region, as shown in Fig.\ref{fig2}(a). 
\par
As a characteristic temperature to describe this non-monotonic behavior of $\chi$, we conveniently define the spin-gap temperature $T^*$ as the temperature at which the spin susceptibility takes a maximum value. When we plot $T^*$ in the BCS-BEC crossover region, we obtain Fig.\ref{fig2}(b). Although $T^*$ is not accompanied by any phase transition, this crossover temperature physically means that preformed singlet pairs start to appear around this temperature.  
\par
\begin{figure}
\begin{center}
\includegraphics[width=10cm]{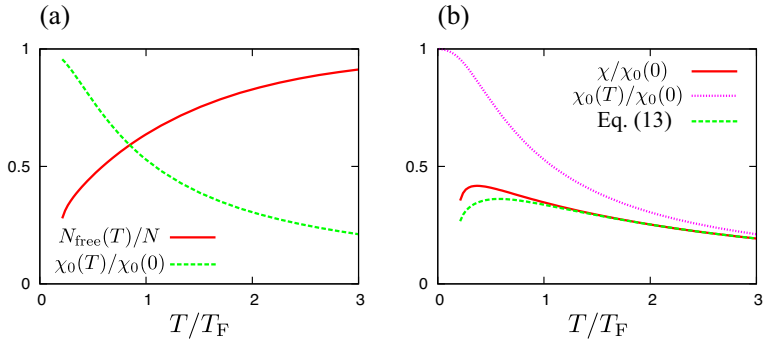}
\end{center}
\caption{(Color online) (a) The number of free Fermi atoms $N_{\rm free}$, and spin susceptibility $\chi_0(T)$ for a free Fermi gas, as functions of temperature. (b) Comparison of $\chi$ calculated in the extended $T$-matrix approximation and Eq. (\ref{eq.100}). In this figure, we take $(k_{\rm F}a_{s})^{-1}=0$.}
\label{fig3}
\end{figure}

To understand the non-monotonic behavior of $\chi$ in a simple manner, we divide the number equation (\ref{eq.9}) into the sum of the non-interacting part,
\begin{equation}
N_{\rm free}= T\sum_{\mbf{p},i\omega_{n}}\sum_{\sigma}G_{0\sigma}(\mbf{p},i\omega_{n}),
\end{equation}
and the fluctuation contribution $N_{\rm fluc}$, as $N=N_{\rm free}+N_{\rm fluc}$. When we simply assume that, while the non-interacting part behaves as a free Fermi gas, the fluctuation component $N_{\rm fluc}$ does not contribute to $\chi$ because of their singlet formation, $\chi$ may be simply estimated as
\begin{equation}
\chi(T)\simeq\chi_0(T)\times{N_{\rm free}(T) \over N}.
\label{eq.100}
\end{equation}
As shown in Fig.\ref{fig3}(a), while $\chi_0(T)$ simply increases with increasing the temperature, $N_{\rm free}$ decreases due to the enhancement of pairing fluctuations near $T_{\rm c}$. As a result, Eq.(\ref{eq.100}) exhibits a non-monotonic temperature dependence, which is qualitatively consistent with $\chi$ in the extended $T$-matrix theory. (See Fig.\ref{fig3}(b).) 
\par
We note that, in contrast to the complicated high-$T_{\rm c}$ cuprates, the BCS-BEC crossover physics of cold Fermi gases is simply dominated by pairing fluctuations. Thus, both the pseudogap phenomenon and spin-gap phenomenon in this system are attributed to fluctuations in the Cooper channel. In this case, if the pseudogap really exists in the single-particle density of states, the `(pseudo)gap energy' would reflect the `binding energy' of a preformed Cooper pair. Thus, the spin-gap temperature $T^*$ is expected to be related to the pseudogap temperature ($T_{\rm pg}$) evaluated from the density of states\cite{stsuchiya09,rwatanabe12}. However, comparing $T^*$ with $T_{\rm pg}$, the former is found to be higher than the latter\cite{stsuchiya09,rwatanabe12}. This indicates that the spin susceptibility is more sensitive to pairing fluctuations being accompanied by preformed pairs than the density of states (although the detailed comparison of $T^*$ with $T_{\rm pg}$ somehow depends on their definitions, because they are crossover temperatures).

%%%%%%%%%%%%%%%%%%%%%%%%%%%%%%%%%%%%%%%%
%%%%%%%%%%%%%%%%%%%%%%%%%%%%%%%%%%%%%%%%

%%%%%%%%%%%%%%%%%%%%%%%%%%%%%%%%%%%%%%%%%%%%%%%%%%%%%%%%%%%%%%%%%%%%%%%%%%%%%%%
\section{Summary}
To summarize, we have discussed the spin susceptibility $\chi$ in the BCS-BEC crossover regime of an ultracold Fermi gas. Within the framework of an extended $T$-matrix theory, we showed that $\chi$ exhibits a non-monotonic temperature dependence in the crossover region, reflecting the formation of preformed singlet Cooper pairs. We have also determined the spin-gap temperature $T^*$ as the temperature at which $\chi$ takes a maximum value. This crossover temperature is higher than the pseudogap temperature evaluated from the single-particle density of states, indicating that the spin susceptibility is a useful probe to examine strong-coupling phenomena in this system. Since the pseudogap phenomenon is an important topic in the BCS-BEC crossover regime of ultracold Fermi gases, our results would be helpful for further understanding of this many-body phenomenon.
\par
%%%%%%%%%%%%%%%%%%%%%%%%%%%%%%%%%%%%%%%%%%%%%%%%%%%%%%%%%%%%%%%%%%%%%%%%%%%%%%
\par
\begin{acknowledgements}
We would like to thank T. Kashimura and D. Inotani for useful discussions. Y.O was supported by Grant-in-Aid for Scientific Research from MEXT in Japan (25400418, 25105511, 23500056).
\end{acknowledgements}

%%%%%%%%%%%%%%%%%%%%%%%%%%%%%%%%%%%%%%%%%%%%%%%%%%%%%%%%%%%%%%%%%%%%%%%%%%%%%%


\begin{thebibliography}{99}
\bibitem{sgiorgini08} S. Giorgini, S. Pitaevskii, S. Stringari, Rev. Mod. Phys. \textbf{80}, 1215 (2008).
\bibitem{ibloch08} I. Bloch, J. Dalibard, W. Zwerger, Rev. Mod. Phys. \textbf{80}, 885 (2008).
\bibitem{cchin10} C. Chin, R. Grimm, P. Julienne, E. Tiesinga, Rev. Mod. Phys. \textbf{82}, 1225 (2010).
\bibitem{yohashi03} Y. Ohashi, A. Griffin, Phys. Rev. A \textbf{67}, 063612 (2003).

\bibitem{stsuchiya09} S. Tsuchiya, R. Watanabe, Y. Ohashi, Phys. Rev. A \textbf{80}, 033613 (2009); \textbf{82}, 033629 (2010); \textbf{84}, 043647 (2011).
\bibitem{Chen2}Q. Chen and K. Levin, Phys. Rev. Lett. \textbf{102}, 190402 (2009).
\bibitem{Magierski}P. Magierski, G. Wlazlowski, A. Bulgac, and J. E. Drut, Phys. Rev. Lett. \textbf{103}, 210403 (2009).
\bibitem{rwatanabe10} R. Watanabe, S. Tsuchiya, Y. Ohashi, Phys. Rev. A \textbf{82}, 043630 (2010); \textbf{85}, 039908(E) (2012).
\bibitem{Hu} H. Hu, X.-J. Liu, P. D. Drummond, and H. Dong, Phys. Rev. Lett. \textbf{104}, 240407 (2010).
\bibitem{Su} S. Su, D. Sheehy, J. Moreno, and M. Jarrell, Phys. Rev. A \textbf{81}, 051604 (2010).
\bibitem{Perali} A. Perali, F. Palestini, P. Pieri, G. C. Strinati, J. T. Stewart, J. P. Gaebler, T. E. Drake, and D. S. Jin, Phys. Rev. Lett. \textbf{106}, 060402 (2011).
\bibitem{Mueller}E. J. Mueller, Phys. Rev. A \textbf{83}, 053623 (2011).
\bibitem{jtstewart08} J. T. Stewart, J.P. Gaebler, D. S. Jin, Nature \textbf{454}, 744 (2008).
\bibitem{jpgaebler10} J. P. Gaebler, J. T. Stewart, T. E. Drake, D. S. Jin, A. Perali, P. Pieri, G. C. Strinati, Nat. Phys. \textbf{6}, 569 (2010).
\bibitem{nascimbene11} S. Nascimb\`{e}ne, N. Navon, S. Pilati, F. Chevy, S. Giorgini, A. Georges, C. Salomon, Phys. Rev. Lett. \textbf{106}, 215303 (2011).
\bibitem{rwatanabe12} R. Watanabe, S. Tsuchiya, Y. Ohashi, Phys. Rev. A \textbf{86}, 063603 (2012).
\bibitem{chrenner98} Ch. Renner, B. Revaz, J.-Y. Genoud, K. Kadowaki, \O. Fischer, Phys. Rev. Lett. \textbf{80}, 149 (1998).
\bibitem{yyoshinari90} Y. Yoshinari, H. Yasuoka, Y. Ueda, K. Koga, K. Kosuge, J. Phys. Soc. Jpn. \textbf{59}, 3698 (1990).
\bibitem{Sanner} C. Sanner, E. J. Su, A. Keshet, W. Huang, J. Gillen, R. Gommers, and W. Ketterle, Phys. Rev. Lett. \textbf{106}, 010402 (2011).
\bibitem{Sommer} A. Sommer, M. Ku, G. Roati, and M. W. Zwierlein, Nature (London) \textbf{472}, 201 (2011).
\bibitem{NSR} P. Nozi\`eres and S. Schmitt-Rink, J. Low Temp. Phys. \textbf{59}, 195 (1985).
\bibitem{SadeMelo} C. A. R. Sa de Melo, M. Randeria, and J. R. Engelbrecht, Phys. Rev. Lett. \textbf{71}, 3202 (1993).
\bibitem{Pieri} P. Pieri and G. C. Strinati, Phys. Rev. B \textbf{61}, 15370 (2000).
\bibitem{Stajic} J. Stajic, J. N. Milstein, Q. Chen, M. L. Chiofalo, M. J. Holland, and K. Levin, Phys. Rev. A \textbf{69}, 063610 (2004).
\bibitem{Liu} X.-J. Liu and H. Hu, Europhys. Lett. \textbf{75}, 364 (2006).
\bibitem{Parish} M. M. Parish, F. M. Marchetti, A. Lamacraft, and B. D. Simons, Nat. Phys. \textbf{3}, 124 (2007).
\bibitem{tkashimura12} T. Kashimura, R. Watanabe, Y. Ohashi, Phys. Rev. A \textbf{86}, 043622 (2012).
\bibitem{Kubo} R. Kubo, {\it Statistical Mechanics} (North-Holland, Amsterdam, 1988) Chap.4.
\end{thebibliography}
\end{document}